\documentclass{emulateapj}

\newcommand{\swift}{{\em Swift}}

\begin{document}

\shorttitle{The Origin of GRB\,090709A}
\shortauthors{Cenko et al.}

\title{Unveiling the Origin of GRB\,090709A: Lack of Periodicity in a
          Reddened Cosmological Long-Duration Gamma-Ray Burst}

\author{S.~B.~Cenko\altaffilmark{1}, N.~R.~Butler\altaffilmark{1,2},
             E.~O.~Ofek\altaffilmark{3,2}, D.~A.~Perley\altaffilmark{1},
             A.~N.~Morgan\altaffilmark{1}, D.~A.~Frail\altaffilmark{4},
             J.~Gorosabel\altaffilmark{5}, J.~S.~Bloom\altaffilmark{1}, 
             A.~J.~Castro-Tirado\altaffilmark{5}, J.~Cepa\altaffilmark{6},
             P.~C.~Chandra\altaffilmark{7}, A.~de Ugarte Postigo\altaffilmark{8},      
             A.~V.~Filippenko\altaffilmark{1},
             C.~R.~Klein\altaffilmark{1}, S.~R.~Kulkarni\altaffilmark{3},
             A.~A.~Miller\altaffilmark{1}, P.~E.~Nugent\altaffilmark{9},
             and D.~L.~Starr\altaffilmark{1,10}}

\altaffiltext{1}{Department of Astronomy, 
  University of California, Berkeley, CA 94720-3411.}
\altaffiltext{2}{Einstein Fellow.}
\altaffiltext{3}{Division of Physics, Mathematics, and Astronomy, California
  Institute of Technology, Pasadena, CA 91125.}
\altaffiltext{4}{National Radio Astronomy Observatory, 1003 Lopezville
  Road, Socorro, NM 87801.}
\altaffiltext{5}{Instituto de Astrof\'{\i}sica de Andaluc\'{\i}a (IAA-CSIC),
  Apartado de Correos, 3.004, E-18.080 Granada, Spain.}
\altaffiltext{6}{Instituto de Astrof\'{i}sica de Canarias, 38205 La Laguna,
  Spain.}
\altaffiltext{7}{Department of Physics, Royal Military College of
  Canada, Kingston, ON, Canada.}
\altaffiltext{8}{INAF/Osservatorio Astronomico di Brera, via Bianchi
  46, 23807 Merate, LC, Italy.}
\altaffiltext{9}{Lawrence Berkeley National Laboratory, 1 Cyclotron Road, 
  Berkeley, CA 94720.}
\altaffiltext{10}{Las Cumbres Observatory Global Telescope Network, 
  Inc., 6740 Cortona Dr. Suite 102, Santa Barbara, CA 93117, USA.}

\slugcomment{Submitted to AJ}

%%%%%%%%%%%%%%%%%%%%%%%%%%%%%%%%%%%%%%%%%%%%%%%%%%%%%%%%%%%%%%%%%%%%%%%%%%%%

\begin{abstract}

We present broadband (gamma-ray, X-ray, near-infrared, optical, and radio) observations 
of the gamma-ray burst (GRB) 090709A and its afterglow in an effort to ascertain the 
origin of this high-energy transient.  Previous analyses suggested that GRB\,090709A
exhibited quasi-periodic oscillations with a period of 8.06\,s, a trait unknown
in long-duration GRBs but typical of flares from soft gamma-ray 
repeaters.  When properly accounting for the underlying shape of the power-density
spectrum of GRB\,090709A, we find no conclusive ($> 3\sigma$) evidence for the
reported periodicity.  In conjunction with the location of the transient (far from
the Galactic plane and absent any nearby host galaxy in the local universe) and 
the evidence for extinction in excess of the Galactic value, we consider a 
magnetar origin relatively
unlikely.  A long-duration GRB, however, can account for the majority 
of the observed properties of this source.  GRB\,090709A is distinguished from
other long-duration GRBs primarily by the large amount of obscuration from 
its host galaxy ($A_{K,\mathrm{obs}} \gtrsim 2$\,mag).
\end{abstract}

%%%%%%%%%%%%%%%%%%%%%%%%%%%%%%%%%%%%%%%%%%%%%%%%%%%%%%%%%%%%%%%%%%%%%%%%%%%%

\keywords{gamma rays: bursts -- stars: neutron}

%%%%%%%%%%%%%%%%%%%%%%%%%%%%%%%%%%%%%%%%%%%%%%%%%%%%%%%%%%%%%%%%%%%%%%%%%%%%

\section{Introduction}
\label{sec:intro}
A variety of astrophysical sources are capable of producing short,
intense flashes of high-energy emission detectable by the current
generation of gamma-ray satellites.  These sources span an 
incredible range of the observable universe, from electrical discharges
associated with thunderstorms on Earth \citep{fbm+94} to the 
deaths of the earliest known stars in the universe \citep{tfl+09,sdc+09}.

Gamma-ray bursts (GRBs) are the most luminous class of these
high-energy transients ($L_{\mathrm{iso}} \approx 10^{50}$--$10^{52}$\,erg\,s$^{-1}$).
At least two distinct progenitor systems are thought to produce
GRBs \citep{kmf+93}.  It is now widely accepted that most
GRBs with duration $t_{90} \gtrsim 2$\,s arise as a byproduct of the 
core collapse of massive stars (hereafter referred to as 
``long-duration'' GRBs; e.g., \citealt{wb06} and references therein).
The origin of ``short-duration'' GRBs is still a hotly debated topic.
They likely arise from an older stellar population
(e.g., \citealt{gso+05,hsg+05,bcb+05,cdg+05,bpc+05,bpp+06,gcg+06}),
possibly due to the merger of a neutron star-neutron star (NS-NS) or 
black hole-neutron star (BH-NS) binary \citep{elp+89,npp92}.

Soft gamma-ray repeaters (SGRs), on the other hand, are distinguished
from GRBs by repeated outbursts with isotropic energy release of 
$E_{\gamma,\mathrm{iso}} \lesssim 10^{46}$\,erg (e.g., \citealt{wt06,o07,m08}).
The discovery of periodic oscillations from bright SGR
flares \citep{mgi+79}, along with the measurement of a spindown 
in their periods \citep{kds+98}, allows for an estimation of the magnetic
field strength.  Unlike typical rotation-powered radio pulsars, 
SGRs (as well as their counterparts
discovered in quiescence, the anomalous X-ray pulsars, or 
AXPs) are likely powered by their intense magnetic fields
($B \gtrsim 10^{14}$\,G; \citealt{dt92}).  Together, SGRs and 
AXPs are now thought to comprise a single class of young, 
highly magnetized neutron stars (``magnetars'').

At least three discoveries have in recent years challenged this simple
classification picture.  First, both GRB\,060614 and GRB\,060505 
had high-energy durations in excess of a few seconds.
Yet despite being quite nearby ($z \approx 0.1$), neither exhibited evidence 
for an associated supernova to quite deep limits
\citep{gnb+06,gfp+06,dcp+06,fwt+06,ocg+07}.  Second,
many GRBs that would have been classified as having short duration ($t_{90}
\lesssim 2$\,s) by the less sensitive \textit{BATSE} satellite exhibit soft 
(and oftentimes faint) X-ray tails
extending hundreds of seconds in time when observed by
\swift\ (e.g., \citealt{bcb+05,pmg+09})
-- perceived duration is, after all, at once a redshift-, sensitivity-, and 
bandpass-dependent quantity.  Finally, 
the X-ray and optical afterglow 
light curves of GRB\,070610 ($t_{90} = 5$\,s) displayed dramatic flares
on time scales as short as several seconds 
\citep{kck+08,sks+08,cdg+08}.  
Though almost certainly of Galactic origin, the source of this variability
is still poorly understood.

Here we present observations of GRB\,090709A, a high-energy transient
whose classification as a traditional long-duration GRB has been 
called into question.  \citet{GCN.9645}
reported the detection of quasi-periodic oscillations (period $P = 
8.06$\,s) in the high-energy light curve, typical of the observed
properties of SGRs and AXPs.  However, unlike all currently known
magnetars, the localization of GRB\,090709A is inconsistent with
both the Galactic plane and any nearby galaxy.  Through a detailed
analysis of the high-energy light curve, the broadband afterglow, 
and the immediate environment, we 
attempt to uncover the origin of GRB\,090709A.

Throughout this work we adopt a standard 
$\Lambda$CDM cosmology ($\Omega_{\Lambda} = 0.73$; 
$\Omega_{m} = 0.27$; $H_{0} = 71$\,km\,s$^{-1}$\,Mpc$^{-1}$).
All quoted uncertainties are 1$\sigma$ (i.e., 68\%) confidence intervals
unless otherwise stated.  Spectral and temporal power-law
indices are provided using the convention $f_{\nu}
\propto t^{-\alpha} \nu^{-\beta}$ \citep{spn98}.  UT dates are used
throughout this work.

In the final stages of preparing this manuscript, \citet{dei+09} posted
an analysis of X-ray (\textit{XMM-Newton} and \textit{Swift}-XRT)
and gamma-ray (\textit{Swift}-BAT and \textit{Integral}-SPI/ACS)
observations of GRB\,090709A.  These authors reach largely similar
conclusions regarding the origin of GRB\,090709A, although they
favor a somewhat larger distance for the event.  We attempt to 
highlight both the differences and similarities between our work 
in what follows.

%%%%%%%%%%%%%%%%%%%%%%%%%%%%%%%%%%%%%%%%%%%%%%%%%%%%%%%%%%%%%%%%%%%%%%%%%%%%

\section{High-Energy Properties}
\label{sec:grb}
\begin{figure}[t!]
  \epsscale{1.3}
  \centerline{\plotone{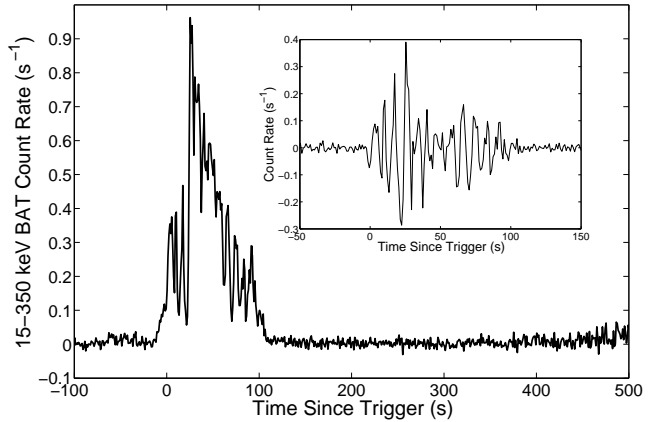}}
  \caption{\swift-BAT 15--350\,keV light curve of GRB\,090709A, 
    referenced to the time of the BAT trigger, 7:38:34 UTC on 9 
    July 2009.  The high-energy emission is dominated by a $\sim 100$\,s long 
    peak beginning at $t_{0}$, although fainter emission is visible both 
    before and after this episode.  The inset shows the residual emission after
    subtracting a smoothed version of the light curve.}
\label{fig:grb}
\end{figure}

GRB\,090709A was detected by the Burst Alert Telescope (BAT: \citealt{bbc+05})
on-board the \swift\ satellite \citep{gcg+04} at 7:38:34 UTC on 9 July 2009
($t_{0}$; \citealt{GCN.9625}).  In Figure~\ref{fig:grb} we plot the 
15--350\,keV BAT light curve,
binned with 1\,s time resolution, obtained following the prescription described 
by \citet{bkb+07}.  The emission is dominated by a broad peak beginning at 
$t_{0}$ and lasting approximately 100\,s. However, there is evidence for 
low-level variability well before the trigger ($t \approx t_{0}-70$\,s), and the 
light curve appears to rise again at $t \approx t_{0}+400$\,s.  Shortly
thereafter the 
spacecraft slewed away due to an observing constraint \citep{GCN.9640}.  Including 
data from $t_{0}-66$\,s to $t_{0}+509$\,s, we measure $t_{90} = 345 \pm 64$\,s 
(15--350\,keV).  Over this interval we find the spectrum
is best fit by a power law with an exponential cutoff, with $\alpha = 
1.06 \pm 0.14$ and $E_{\mathrm{p}} = 299_{-101}^{+547}$\,keV ($\chi^{2} = 
38.71$ for 55 degrees of freedom, d.o.f.).  The corresponding 15--350\,keV fluence is 
$f = 4.59_{-0.26}^{+0.30} \times 10^{-5}$\,erg\,cm$^{-2}$, making GRB\,090709A
one of the brightest \swift\ events detected to date.

Superposed on the overall $\sim 100$\,s duration rise and decline
of the dominant emission component, the
high-energy light curve exhibits relatively strong fluctuations.  In the
inset of Figure~\ref{fig:grb}, we plot the residual emission after subtracting
a smoothed version of the 15--350\,keV light curve (boxcar binned by 10\,s).

%%%%%%%%%%%%%%%%%%%%%%%%%%%%%%%%%%%%%%%%%%%%%%%%%%%%%%%%%%%%%%%%%%%%%%%

%%%%%%%%%%%%%%%%%%%%%%%%%%%%%%%%%%%%%%%%%%%%%%%%%%%%%%%%%%%%%%%%%%%%%%%%%%%  

\citet{GCN.9645} first reported a search for a periodic signal from 
GRB\,090709A, claiming a detection of an excess in the
power density spectrum (PDS) at $P = 8.06$\,s.  To estimate the
significance of this excess, \citet{GCN.9645} normalize the entire PDS 
by the noise in the observed frequency range 0.2--0.6\,Hz.  After 
correcting for the number of frequency bins examined ($\sim 2000$) and
the estimated number of bursts for which such an analysis could be 
performed ($\sim 100$), the authors conclude the observed peak is 
highly significant (null probability $\sim 10^{-6}$). 
The detection of this apparent periodicity was subsequently confirmed on several
additional high-energy satellites 
\citep{GCN.9647,GCN.9649,GCN.9653}.  

\begin{figure*}[t!]
  \epsscale{1.2}
  \centerline{\plottwo{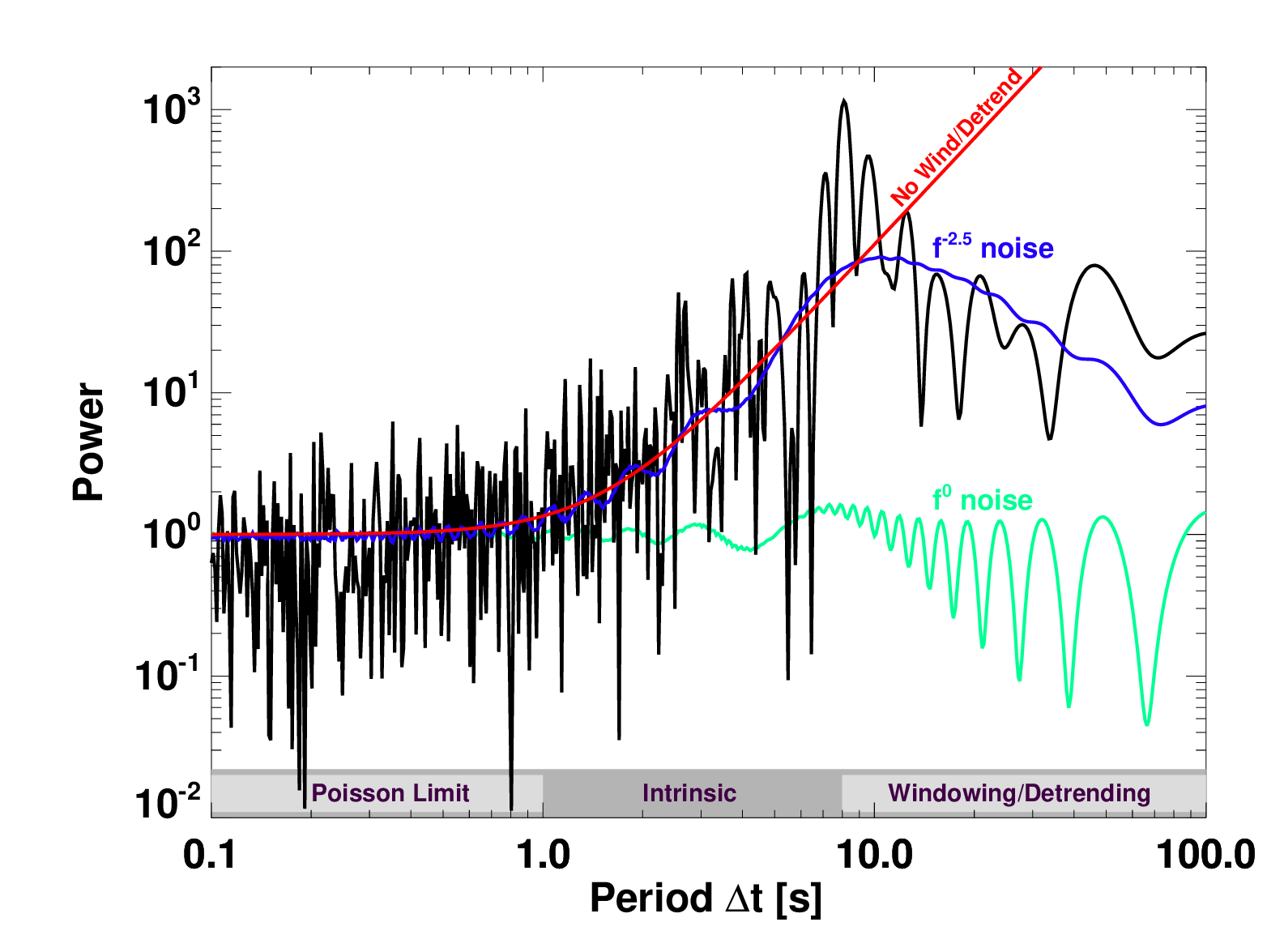}{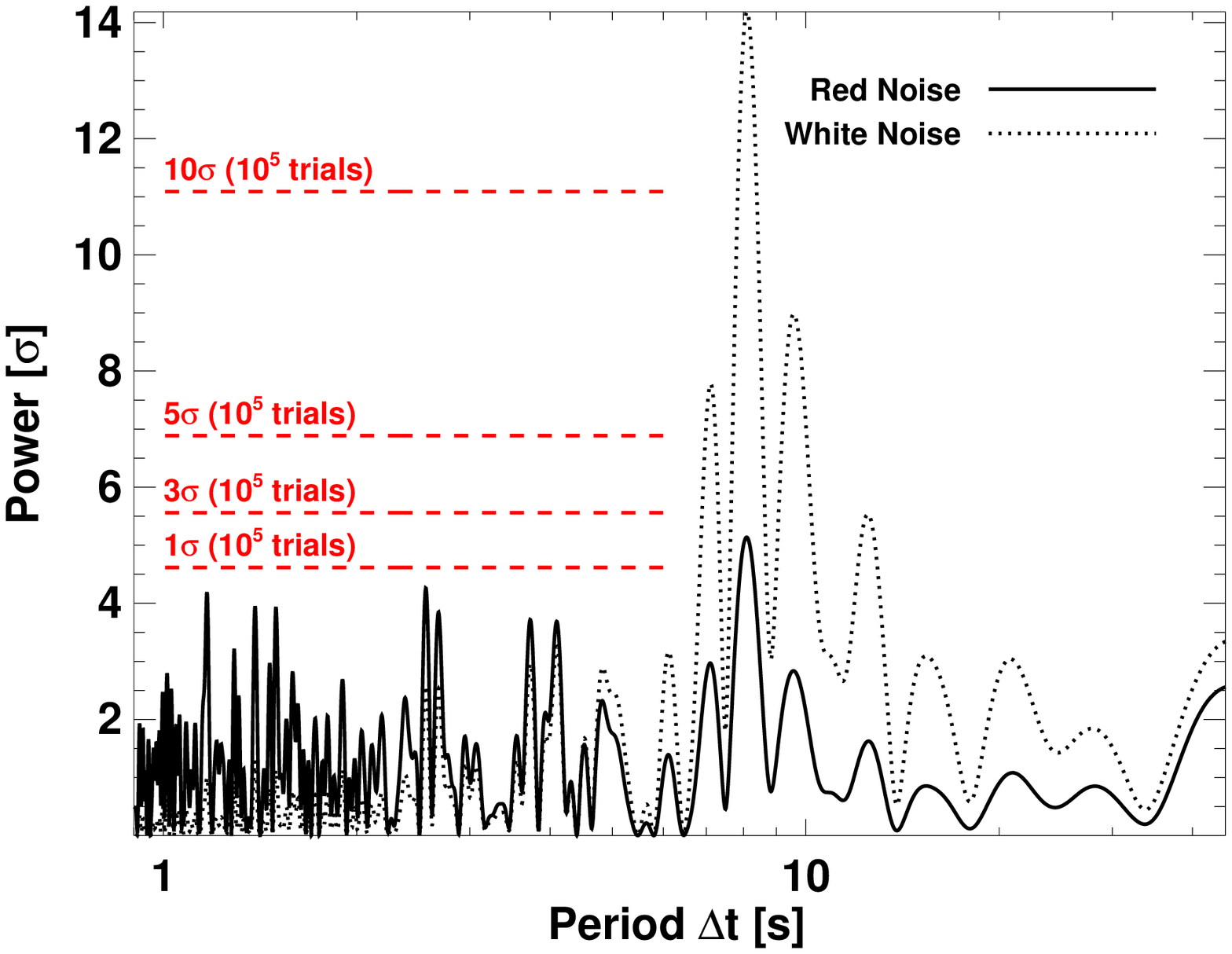}}
  \caption{\textit{Left:} Un-normalized PDS of GRB\,090709A.  The PDS was
                 constructed using the 15--350\,keV BAT light curve from $t_{0}$
                 to $t_{0} + 100$\,s.  A smoothed (10\,s boxcar) version of the light
                 curve was first subtracted to remove the overall rise and decline
                 trend.  Three clear regions are visible.  For $P \lesssim 1$\,s, the
                 PDS is dominated by Poisson statistics (i.e., white noise).  For
                 $P \gtrsim 10$\,s, the effects of windowing and de-trending
                 dominate the error in the PDS.  The power-law slope in the
                 intermediate regime ($\alpha = 2.5$), intrinsic to the GRB prompt
                 emission but simulating the effects of red noise, is clearly 
                 visible.  The red line shows our model for the underlying shape
                 of the PDS derived using a non-windowed light curve absent
                 de-trending.  Noise models assuming a flat PDS (green line, 
                 $f^{0}$ noise) and a power-law PDS (blue line, $f^{2.5}$ noise)
                 are also shown.  \textit{Right:} The normalized PDS of 
                 GRB\,090709A.  The dashed line assumes a flat underlying
                 spectrum (white noise), while the solid line accounts
                 for the intrinsic fluctuations (red noise).  Significance intervals
                 assuming 10$^{5}$ trials are indicated by the horizontal dashed
                 blue lines.  The significance of the observed feature at 
                 $P = 8.1$\,s drops dramatically after accounting for the
                 underlying power-law spectrum.}
\label{fig:pds}
\end{figure*}

In the left panel of Figure~\ref{fig:pds} we plot the un-normalized
PDS of GRB\,090709A.  The 15--350\,keV light curve from $t_{0}$
to $t_{0}+100$\,s was binned with a time resolution of 10\,ms, and 
de-trended by subtracting a smoothed version of the light
curve (10\,s boxcar).  The peak noted by \citet{GCN.9645} is clearly
visible at a period of 8.1\,s.  It is crucial to note, however, that
\textit{the interpretation of the significance of this peak depends 
sensitively on the assumed noise properties and the underlying shape
of the PDS}.  We therefore examine this issue here in greater detail.

The simplest strategy to infer the statistical significance of this feature
is to assume that the PDS is dominated by white noise (i.e., independent 
of frequency).  The significance (signal-to-noise ratio, SNR) can then be 
estimated by normalizing the PDS by the observed scatter in a region
devoid of features.  Following \citet{GCN.9645}, we normalize the PDS
of GRB\,090709A with respect to the observed scatter in the range
$P = 3$--6.5\,s (one of the noisier regions of the PDS).  The result
is shown as the dashed line in the right panel of Figure~\ref{fig:pds}.
After correcting for $\sim 10^{5}$ trials ($\sim 1000$ period bins 
using our 10\,ms light curve and considering $0 \lesssim P \lesssim
10$\,s for $\sim 100$ long-duration GRBs), we find that the observed
peak at $P = 8.1$\,s is significant at the $12\sigma$ level.

%%%%%%%%%%%%%%%%%%%%%%%%%%%%%%%%%%%%%%%%%%%%%%%%%%%%%%%%%%%%%%%
%%%%%%%%%%%%%%%%%%%%%%%%%%%%%%%%%%%%%%%%%%%%%%%%%%%%%%%%%%%%%%%

The PDSs of GRB prompt emission, however, are not featureless. 
Examining the left panel of Figure~\ref{fig:pds}, three distinct
regions can be defined.  For the shortest periods ($P \lesssim 1$\,s),
the PDS is relatively flat and dominated by Poisson statistical fluctuations
(white noise).  For large periods ($P \gtrsim 10$\,s), the PDS
is dominated by the total duration of the analyzed light curve (the
window function, $\sim 100$\,s for GRB\,090709A) and the
de-trending algorithm ($\sim 10$\,s smoothing for GRB\,090709A).

For intermediate periods ($1 \lesssim P \lesssim 10$\,s), 
\citet{bss98} have shown that the PDSs of the longest ($t_{90}
> 100$\,s) \textit{BATSE} bursts are well-fit by a power law
with index $\alpha \approx 5/3$ (where power $\propto P^{\alpha}$).  
Since the average
power-law index was found to be in close agreement with the
Kolmogorov law, it has been suggested that the prompt emission
may be related to turbulence in the outflow \citep{bss00}.  Subsequent
analysis has indicated, however, that the exact value of the power-law
slope can vary significantly from event to event, and is determined
predominantly by the shape and duration of individual pulse episodes 
within the GRB \citep{smj02}.  PDS measurements of bright \swift\
events have been performed (e.g., GRB\,080319B, \citealt{bpl+09})
and show results consistent with those from \textit{BATSE}.

Regardless of its origin, it is clear that we must include the underlying
behavior of the PDS when evaluating the significance of individual features.
The underlying slope is not ``noise'' in the sense that it is not caused by
the limitations of the measuring instrument; the observed power-law
PDS is an intrinsic property of the GRB itself.  Nonetheless, we can remove this 
contribution much as we would eliminate red ($\alpha = 2$) or pink
($0 < \alpha < 2$) noise caused by our measuring device 
(e.g., \citealt{v05}).  

To estimate the power-law slope, we construct a new PDS for
GRB\,090709A, including all of the available 15--350\,keV BAT 
data and without any de-trending.  This helps to remove the
noise at $P \gtrsim 10$\,s and provides a longer lever arm to
calculate the PDS slope.  For GRB\,090709A, we then find
$\alpha \approx 2.5$.

In the right panel of Figure~\ref{fig:pds}, we re-normalize the PDS of 
GRB\,090709A using the underlying spectrum described above
(solid line).  While the same shape is visible in the PDS, the
significance of the peak has dropped dramatically.  After correcting for the number
of trials, we find that the feature at $P = 8.1$\,s is significant at only the
$\sim 2\sigma$ level.

\begin{figure*}[t!]
  \centerline{\plotone{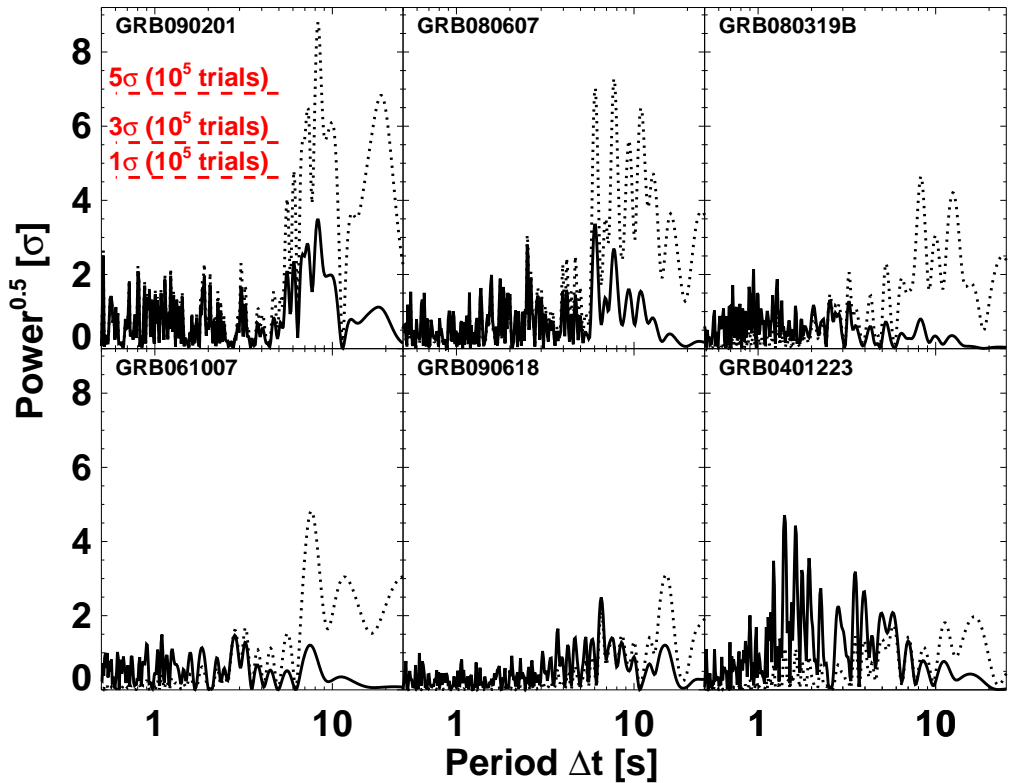}}
  \caption{The PDSs of other \swift\ long-duration GRBs.  We have selected all BAT events
                 with fluence $> 70\%$ that of GRB\,090709A and duration $t_{90} > 70$\,s.  As
                 in the right panel of Figure~\ref{fig:pds}, the dashed lines plot the 
                 normalized PDS assuming a flat underlying spectrum using the noise
                 properties in the range $3 < P < 6.5$\,s.  All six events have a peak at $P \approx
                 10$\,s with single-trial significance $> 2\sigma$, though none is as 
                 strong as that derived for GRB\,090709A.  After correcting for the 
                 underlying power-law spectrum (solid lines) and the number of trials, none 
                 of these features remain with significance $> 3\sigma$. }
\label{fig:others}
\end{figure*}

As stated previously, the inferred significance depends sensitively on the 
assumed noise properties.  The value we derive for 
GRB\,090709A ($\alpha = 2.5$) is significantly steeper than the average
\textit{BATSE} value of $\sim 5/3$ \citep{bss98}.  To investigate further,
we have performed a comparable analysis on all BAT GRBs with a fluence
at least 70\% of the value derived for GRB\,090709A (to provide sufficient
SNR) and a duration $t_{90} > 70$\,s (to provide sufficient sensitivity
to $P \approx 10$\,s).  Only 6 additional events meet this criteria: 
GRBs\,041223, 061007, 080319B, 080607, 090201, and 090618.  The results
of this analysis are shown in Table~\ref{tab:others} and 
Figure~\ref{fig:others}.

%%%%%%%%%%%%%%%%%%%%%%%%%%%%%%%%%%%%%%%%%%%%%%%%%%%%%%%%%%%%%%%%%%%%%%%
\input{others.tab}
%%%%%%%%%%%%%%%%%%%%%%%%%%%%%%%%%%%%%%%%%%%%%%%%%%%%%%%%%%%%%%%%%%%%%%%

Each event in this sample exhibits a peak in the PDS with period $\sim 10$\,s
and significance greater than 2$\sigma$ when normalizing by a flat PDS.
However, the significance of these features drops dramatically when normalizing
by a power-law spectrum with index $\alpha \approx 2$--2.5.  

The apparent periodicity of GRB\,090709A was also reported by 
several additional gamma-ray
satellites \citep{GCN.9647,GCN.9649,GCN.9653}, presumably performing a 
similar analysis to that of \citet{GCN.9645}.  We repeated our analysis
on the publicly available \textit{Suzaku} light curve of GRB\,090709A,
and find our results essentially unchanged: after normalizing by a power law
with index $\alpha \approx 2.5$, the significance of the feature at $P=8.1$\,s
falls to $\sim 2$--3$\sigma$.  This is not entirely surprising, though, 
as the PDS power-law index has been shown to be relatively independent
of bandpass for a given GRB \citep{bss98}.

\citet{dei+09} have performed a similar timing analysis of both the 
\swift-BAT and \textit{Integral}-API/ACS \citep{GCN.9649}
light curves of GRB\,090709A.  These authors confirm our primary result
that the reported periodicity in the BAT light curve at $P = 8.1$\,s is only 
significant at the $\lesssim 3\sigma$ level.  Interestingly, the signal
at this frequency appears to be much weaker in the SPI-ACS data. 

To summarize, while the high-energy light curve of GRB\,090709A shows tantalizing
evidence of quasi-periodic oscillations with $P = 8.1$\,s, we do not
take it to be a significant feature required by the data.  It is certainly
the most compelling candidate for such behavior amongst the long-duration
GRB population observed by \swift.  However, when properly accounting
for the shape of the underlying spectrum, we find the significance of the 
observed periodicity is not sufficiently large to be conclusive.  Instead,
we must look to the afterglow and environment of GRB\,090709A to
attempt to unveil its progenitor.

%%%%%%%%%%%%%%%%%%%%%%%%%%%%%%%%%%%%%%%%%%%%%%%%%%%%%%%%%%%%%%%
%%%%%%%%%%%%%%%%%%%%%%%%%%%%%%%%%%%%%%%%%%%%%%%%%%%%%%%%%%%%%%%

%%%%%%%%%%%%%%%%%%%%%%%%%%%%%%%%%%%%%%%%%%%%%%%%%%%%%%%%%%%%%%%%%%%%%%%%

\section{Follow-Up Observations}
\label{sec:obs}
The X-Ray Telescope (XRT; \citealt{bhn+05}) onboard \swift\ began 
observing the field of GRB\,090709A beginning at $t_{0} + 74$\,s.  A fading
X-ray counterpart inside the BAT error circle was promptly identified 
and reported to the community
\citep{GCN.9625,GCN.9636,GCN.9642}.  Cross-matching field sources 
detected by the XRT with cataloged near-infrared (NIR) positions from
2MASS \citep{scs+06}, we measure a localization for the afterglow of
$\alpha = 19^{\mathrm{h}} 19^{\mathrm{m}} 42\farcs46$, $\delta = +60^{\circ}
43\arcmin 39\farcs6$, with a 90\% containment radius of $1\farcs2$
(J2000.0; see \citealt{b07} for details).
   
We plot the X-ray afterglow of GRB\,090709A in Figure~\ref{fig:aglow}, which 
we have taken from the online compilation of 
N.R.B.\footnote{http://astro.berkeley.edu/$\sim$nat/swift; see \citet{bk07} 
for details.}  Like many GRBs in the \swift\ era, the XRT began observations 
of GRB\,090709A while the prompt gamma-ray emission was still ongoing, and the earliest 
X-ray observations extrapolate smoothly to the tail of the prompt emission.
After $10^{3}$\,s, the X-ray light curve is well described by a 
single power-law decay with index $\alpha_{X} = 1.38 \pm 0.02$ ($\chi^{2} = 
394.93$ for 417 d.o.f.).  However, there is a large gap in the X-ray coverage
from $t \approx 2$\,days to $t \approx 10$\,days.  A broken power law with
$2 \lesssim t_{b} \lesssim 10$\,days can also provide a reasonable fit.
In this case, the initial decay index is somewhat more shallow 
($\alpha_{1,X} \approx 1.2$); the break time and post-break decay index
are not well constrained.  

The X-ray spectrum at this epoch is adequately fit 
by a power law with index $\beta_{X} = 0.95 \pm 0.07$ ($\chi^{2} = 178.58$ for 
175 d.o.f.), although the inferred X-ray column ($N_{\rm H} = 1.83^{+0.24}_{-0.21} 
\times 10^{21}$\,cm$^{-2}$ at $z = 0$) is significantly in excess of the 
Galactic value ($N_{\rm H,Gal} = 6.6 \times 10^{20}$\,cm$^{-2}$; 
\citealt{kbh+05}).  In the case of an intrinsic power-law spectrum,
the requirement for extinction in excess of the Galactic 
value is significant at the 13.5$\sigma$ level.  Fits with only
Galactic extinction require at least two blackbody components
and still provide worse quality than an absorbed power law
($kT_{1} = 0.32 \pm 0.02$\,keV, $kT_{2} = 1.11 \pm 0.05$\,keV;
$\chi^{2} = 198.64$ for 174 d.o.f.).
 These results are broadly consistent with those reported by 
\citet{GCN.9642}.

Both the Peters Automated Infrared Telescope (PAIRITEL; \citealt{bsb+06}) 
and the robotic Palomar 60\,inch (1.5\,m) telescope (P60; \citealt{cfm+06}) 
automatically responded to the \swift\ alert and began observations within 
2 minutes of the trigger time.  Additional observations were obtained at 
later times with the OMEGA$_{2000}$ NIR camera on the 3.5\,m
telescope at Calar Alto Observatory (10 July 2009), with the OSIRIS
instrument (Cepa et al.~2009, in preparation) attached to the 10.4\,m
Gran Telescopio Canarias (GTC) at the IAC's Observatorio del Roque de los 
Muchachos in La Palma (11 July 2009), and with NIRC2 behind the 
laser guide star adaptive optics system \citep{wlb+06} on the 10\,m Keck II 
telescope (17 July 2009).  All images were processed using standard routines 
in the IRAF\footnote{IRAF is distributed by the National Optical Astronomy 
Observatory, which is operated by the Association for Research in Astronomy, 
Inc., under cooperative agreement with the National Science Foundation.} 
environment.  Photometric calibration was performed relative to the 
USNO-B1.0 catalog \citep{mlc+03} in the optical, using the filter transformations 
of \citet{jga06} where appropriate, and relative to 2MASS \citep{scs+06} 
in the NIR.  The results of this campaign are listed in Table~\ref{tab:obs} 
and displayed in Figure~\ref{fig:aglow}.

\begin{figure*}[t]
  \epsscale{1.1}
  \centerline{\plotone{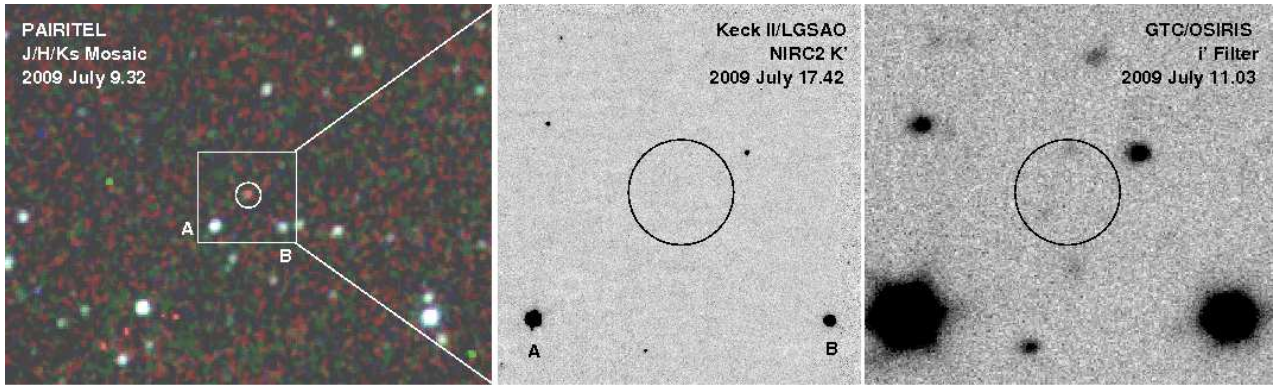}}
  \caption{\textit{Left:} PAIRITEL false-color ($JHK_{s}$) image of the
   field of GRB\,090709A, taken at a mean epoch of $t_{0} + 379$\,s. 
   The afterglow is centered inside the white circle (5\arcsec\ radius),
   and is clearly much redder than the surrounding field stars.  Stars
   $A$ and $B$ are marked for reference in all images.
   \textit{Center:} Keck/NIRC2 $K^{\prime}$ image of the same field,
   approximately one week later.  The black circle has a radius
   of 5\arcsec.  \textit{Right:} GTC/OSIRIS $i^{\prime}$ image of the identical
   field on 11 July 2009.  No sources are visible in the 
   immediate environment ($\lesssim 4\arcsec$) of the afterglow location to a 3$\sigma$
   limiting magnitude of $K^{\prime} > 22.8$\,mag, $i^{\prime} > 25.5$\,mag.  All 
   images are oriented with north up and east to the left.}
\label{fig:finder}
\end{figure*}

\citet{GCN.9635} reported the PAIRITEL afterglow detection in the 
$H$ and $K_{s}$ filters; subsequent re-analysis has revealed a detection 
in the $J$ band as well (Figure~\ref{fig:finder}).  Using early-time $K_{s}$-band
images and reference objects from the 2MASS catalog, we measure a 
position for the afterglow of $\alpha = 19^{\mathrm{h}} 19^{\mathrm{m}} 
42\farcs64$, $\delta = +60^{\circ} 43\arcmin 39\farcs3$, with a 90\%
containment radius of $0\farcs4$ (J2000.0).  This location is at a distance of
$1\farcs4$ from the center of the X-ray error circle, consistent 
within the 90\% confidence regions.

Combining the PAIRITEL NIR detections with the
marginal $r^{\prime}$ and $z^{\prime}$ detections from P60, we find that the 
afterglow was extremely red.  After correcting for Galactic extinction 
($E[B-V] = 0.09$\,mag; \citealt{sfd98}), we fit the nearly simultaneous
$r^{\prime}z^{\prime}JHK_{s}$ photometry at $t \approx 350$\,s to
a power-law spectral energy distribution (SED) model and find an
extremely steep spectral index: $\beta_{O} = 3.8 \pm 0.3$ ($\chi^{2} = 4.16$
for 3 d.o.f.).  The temporal decay index in the optical is not very well
constrained by our observations; assuming the same value for all 
filters, we find $\alpha_{O} \approx 0.9$.

We observed the field of GRB\,090709A with the Very Large Array 
(VLA)\footnote{The Very Large Array is operated by the National Radio 
Astronomy Observatory, a facility of the National Science Foundation 
operated under cooperative agreement by Associated Universities, Inc.} 
at a frequency of 8.46\,GHz on July 11.42 and at 1.43\,GHz on July 11.39.  
For both observations the array was in the ``C'' configuration.  The 
flux-density scale was tied to 3C\,286 or 3C\,147 and the phase was measured 
by switching between the GRB and a nearby, bright, point-source calibrator.  To 
maximize sensitivity, the full VLA continuum bandwidth (100 MHz) was 
recorded in two 50 MHz bands.  Data reduction was carried out following 
standard practice in the AIPS software package.  No emission is 
detected at the NIR afterglow location at either frequency to 2$\sigma$ limits 
of $f_{\nu}$(8.46\,GHz) $< 70$\,$\mu$Jy, $f_{\nu}$(1.43\,GHz) $< 
288$\,$\mu$Jy (see also \citealt{GCN.9657}).

Finally, we have obtained a series of pre-GRB optical images with the 
Palomar 48\,inch (1.2\,m) telescope from the Deep Sky catalog\footnote{See 
http://supernova.lbl.gov/$\sim$nugent/deepsky.html.}.  32 images of the field of 
GRB\,090709A were taken as part of the Palomar-QUEST survey \citep{dbm+08}
over a time interval of a month in August 2005.  The images were obtained using
an order-blocking filter with a cutoff blueward of $\lambda \approx 
6100$\,\AA, which we calibrate relative to the Sloan Digital Sky Survey
$i^{\prime}$ band.  A coadded image of these frames does not reveal any 
source at the location of the afterglow of GRB\,090709A to a 
3$\sigma$ limit of $i^{\prime} > 23.2$\,mag. 

%%%%%%%%%%%%%%%%%%%%%%%%%%%%%%%%%%%%%%%%%%%%%%%%%%%%%%%%%%%%%%%%%%%%%%%%%%%%

%%%%%%%%%%%%%%%%%%%%%%%%%%%%%%%%%%%%%%%%%%%%%%%%%%%%%%%%%%%%%%%%%%%%%%%%%%%%%

%%%%%%%%%%%%%%%%%%%%%%%%%%%%%%%%%%%%%%%%%%%%%%%%%%%%%%%%%%%%%%%%%%%%%%%%%%%

\section{A New Soft Gamma-Ray Repeater?}
\label{sec:magnetar}
The detection of quasi-periodic oscillations with
$P = 8.1$\,s would strongly suggest an SGR origin.  Unlike normal
(rotation-powered) pulsars, the spin periods of known SGRs and
AXPs fall in the range 2.0--11.8\,s
\footnote{See, for example, the McGill SGR/AXP Online Catalog
at http://www.physics.mcgill.ca/$\sim$pulsar/magnetar/main.html.}.
We therefore consider if the observed properties (outside the prompt
high-energy emission) could themselves be consistent with an SGR flare.

By definition SGRs undergo repeated high-energy outbursts.
We have therefore searched the historical GRB catalogs of both the 
Interplanetary Network (IPN: \citealt{lbh+97,hbk+99}) and 
\textit{BATSE} for an excess of 
localizations consistent with the position of GRB\,090709A (both missions 
provided relatively coarse angular localizations, requiring such a statistical
analysis; see \citealt{o07} for details).  Using the IPN catalog from
12 November 1990 to 31 October 2005\footnote{See
http://www.ssl.berkeley.edu/ipn3/interpla.html; version 17 December
2005.  For reference, this IPN catalog version is also available
at http://astro.caltech.edu/$\sim$eran/GRB/IPN/NearbyGal/CatIPN.txt.ver17122005.},
we found four IPN localizations consistent with the position of GRB\,090709A.
However, running the same search on random pointings with the same
ecliptic latitude results in a median overlap of four GRBs.  A similar analysis
with short-duration ($t_{90} < 1$\,s) \textit{BATSE} GRBs in the current
catalog\footnote{See http://gammaray.gsfc.nasa.gov/batse/grb/catalog/current.}
also reveals no significant excess in the direction of GRB\,090709A.

SGR flares are broadly divided into three classes (e.g., \citealt{wt06,m08}).  The rarest
of these are the giant flares, with only a handful observed to date
(e.g., \citealt{mgi+79,hcm+99,ffc+99,mca+99,hbs+05}).  The high-energy light 
curves of giant flares are dominated by a short ($\sim 1$\,s), hard spike of 
gamma-rays.  Thereafter, the emission softens and decays exponentially 
with quasi-periodic oscillations at the spin period of the underlying neutron star.  
Clearly the 
bright initial spike was not observed from GRB\,090709A.  However it is 
possible that this component could be non-isotropic 
and beamed away from our line of sight, as some bright SGR flares lacking 
the initial spike have been observed (e.g., \citealt{mte+09}).

Giant SGR flares have gamma-ray energy releases $\sim 10^{44}$--$10^{46}$\,erg.
Using our derived 15--350\,keV fluence (\S~\ref{sec:grb}), the observed system 
would fall at a distance $\sim 0.1$--$1$\,Mpc (note that any beaming correction
would only make the event more nearby, a possibility we discuss subsequently).
Such distances imply an origin either in the halo of the Milky Way or in a nearby
galaxy in the Local Group.

Magnetars have relatively short characteristic ages ($P / \dot{P} \sim 
10^{3}$--$10^{5}$\,yr), which would seem to rule out a halo origin if the 
neutron star formed directly from the core collapse of a massive star.
If the progenitor system were associated with an older stellar population,
for instance the accretion-induced collapse (AIC) of a white dwarf
\citep{cs76}, this could perhaps explain the large offset from a site of
recent star formation.  However, a relatively unphysical model would be required
to account for the observed X-ray spectrum without excess extinction 
(see below).

Giant magnetar flares from nearby galaxies are now believed to account for 
some fraction of the observed \textit{BATSE} short bursts (e.g., 
\citealt{hbs+05,gkg+05,ccr+05,tcl+05,okn+06,o07,omq+08,mac+08,hbp+09,cpt09}).
In the case of GRB\,090709A, however, the lack of an obvious host candidate
seems to disfavor an extragalactic (but nearby; $d \lesssim 1$\,Mpc) origin.  
Our pre- (Deep Sky) and post- (NIRC2) GRB
limits at the location of GRB\,090709A correspond to absolute magnitudes of
$M_{i^{\prime}} \mathrm{(AB)} \gtrsim -2$\,mag, $M_{K^{\prime}} 
\mathrm{(Vega)} \gtrsim -2$\,mag 
for $d \approx 1$\,Mpc, sufficient to detect individual supergiant 
stars, let alone dwarf galaxies (modulo extinction).  
The post-GRB $i^{\prime}$ limits from the GTC
provide even tighter constraints.  

%%%%%%%%%%%%%%%%%%%%%%%%%%%%%%%%%%%%%%%%%%%%%%%%%%%%%%%%%%%%%%%%%%%
\input{obs.tab}
%%%%%%%%%%%%%%%%%%%%%%%%%%%%%%%%%%%%%%%%%%%%%%%%%%%%%%%%%%%%%%%%%%%

Even with a large natal kick velocity 
($10^{4}$\,km\,s$^{-1}$), the short lifetime limits the distance a magnetar 
can travel away from the host-galaxy disk (where it was presumably formed) to 
$\lesssim 1$\,kpc.  For $d \sim 1$\,Mpc,
this corresponds to an angular offset of $\sim 3$\arcmin.  No Local
Group galaxy is known within 10$^{\circ}$ of GRB\,090709A.  The lack of 
any observed candidate host galaxy at this location seems to independently
rule out a giant magnetar flare having extragalactic origin.

Intermediate magnetar flares have durations of order 1\,s or longer and
energy releases $\sim 10^{41}$--$10^{43}$\,erg.  With the observed fluence
of GRB\,090709A, this corresponds to distances of $\sim 4$--40\,kpc.
The primary drawback with this scenario is the location of GRB\,090709A
with respect to the Galactic plane.  With Galactic coordinates ($l, b$) = 
(91.8$^{\circ}$, 20.1$^{\circ}$), a total distance of $\gtrsim 4$\,kpc 
corresponds to a height of $\gtrsim 1$\,kpc above the Galactic plane. 
Of the 18 known and suspected magnetars, only two have Galactic
latitudes $|b| > 2^\circ$: SGR\,0525-66 (in the LMC; 
\citealt{mgg+82,kkm+03,khg+04}), and 
CXOU\,J010043.1-721134 (in the SMC; \citealt{lfm+02})\footnote{The 
localization of an additional SGR candidate, SGR\,0418+5729,
is centered at $b = 5.1^{\circ}$, but the error circle has a radius of 
$\sim 7^{\circ}$ \citep{GCN.9499}.}.
For those SGRs in our Galaxy with known distances, the typical
scale height is $\sim 100$\,pc, as would be expected from their 
young ages.  Furthermore, several are associated with SN remnants
\citep{khg+04,gg07,ggv99,fg81,vg97} or young stellar clusters 
\citep{vhl+00,ce04,mcc+06}, so even in the absence of a
planar birth, some remnant of recent star formation should be evident
nearby.  

Lastly, we consider SGR flares with energy release $< 10^{41}$\,erg.
Such flares are quite common, though typically observed with durations
$\sim 100$\,ms.  The spectra of these flares are well described by
an optically thin thermal bremsstrahlung model, with 
$kT \sim 20$--40\,keV, unlike the observed properties of 
GRB\,090709A.  If we require a scale height of $\lesssim 100$\,pc,
this limits the total distance to GRB\,090709A to be $\lesssim 
300$\,pc.  The energy release would therefore be
$\lesssim 5 \times 10^{38}$\,erg, consistent with observations of low-level
flaring activity from known SGRs and AXPs.  However, the small 
distance would present additional
difficulties.  Nearly all sufficiently well-localized 
magnetars exhibit quiescent X-ray emission at a luminosity of 
$10^{33}$--$10^{36}$\,erg\,s$^{-1}$.  At $d \lesssim 300$\,pc, such
emission should have been detected by X-ray surveys
such as the \textit{ROSAT} All-Sky Survey \citep{vab+99}, 
yet no cataloged X-ray sources are consistent with this position.

Finally, we note that all Galactic models, even if associated with an 
older population and not subject to the requirement of nearby star formation,
would struggle to explain the excess extinction inferred from the X-ray
spectrum and optical SED.  While the intrinsic X-ray and optical
spectra need not be power laws, we find no physically-motivated 
models capable of reproducing the observed X-ray spectrum
and optical SED absent an excess absorption component.

%%%%%%%%%%%%%%%%%%%%%%%%%%%%%%%%%%%%%%%%%%%%%%%%%%%%%%%%%%%%%%%%%%%%%%%%%%%

\section{Massive Star Core Collapse?}
\label{sec:collapsar}
Setting aside again the high-energy light curve, we consider the long-lived
afterglow emission associated with GRB\,090709A.  Radio afterglow emission has
been observed from two giant SGR flares \citep{fkb99,ccr+05,gkg+05}, and fading 
X-ray counterparts are relatively common, even for intermediate flares
(e.g., \citealt{m08}).  However, the physical mechanism underlying this emission is 
thought to be quite different from the process powering long-duration GRB 
afterglows.

In the case of magnetar flares, afterglow emission results from a rapid injection of
relativistic particles into the circumburst medium.  The particles expand 
adiabatically, and, consequently, the light curves decline quite rapidly with time
(power-law indices $\alpha \gtrsim 2.5$; \citealt{fkb99,cw03,ccr+05,gkg+05}).  
For GRB afterglows, a relativistic shock wave accelerates electrons in the 
circumburst medium, causing them to emit synchrotron radiation (e.g., \citealt{p05}).  
As the shock expands outward, it continues to excite electrons further and 
further away from the explosion site, resulting in additional emission from 
larger radii and a slower fading rate ($\alpha \approx 1$--2; \citealt{spn98}).  

A detailed comparison of well-sampled, broadband light curves may therefore be
able to distinguish between these two scenarios.  In the case of GRB\,090709A,
however, the lack of optical/NIR and radio data make such a comparison
difficult.  We therefore attempt to answer a simpler question: is the observed
emission consistent with standard GRB afterglow models?

%%%%%%%%%%%%%%%%%%%%%%%%%%%%%%%%%%%%%%%%%%%%%%%%%%%%%%%%%%%%%%%%%%%%%%

\begin{figure*}[t]
  \epsscale{1.1}
  \centerline{\plottwo{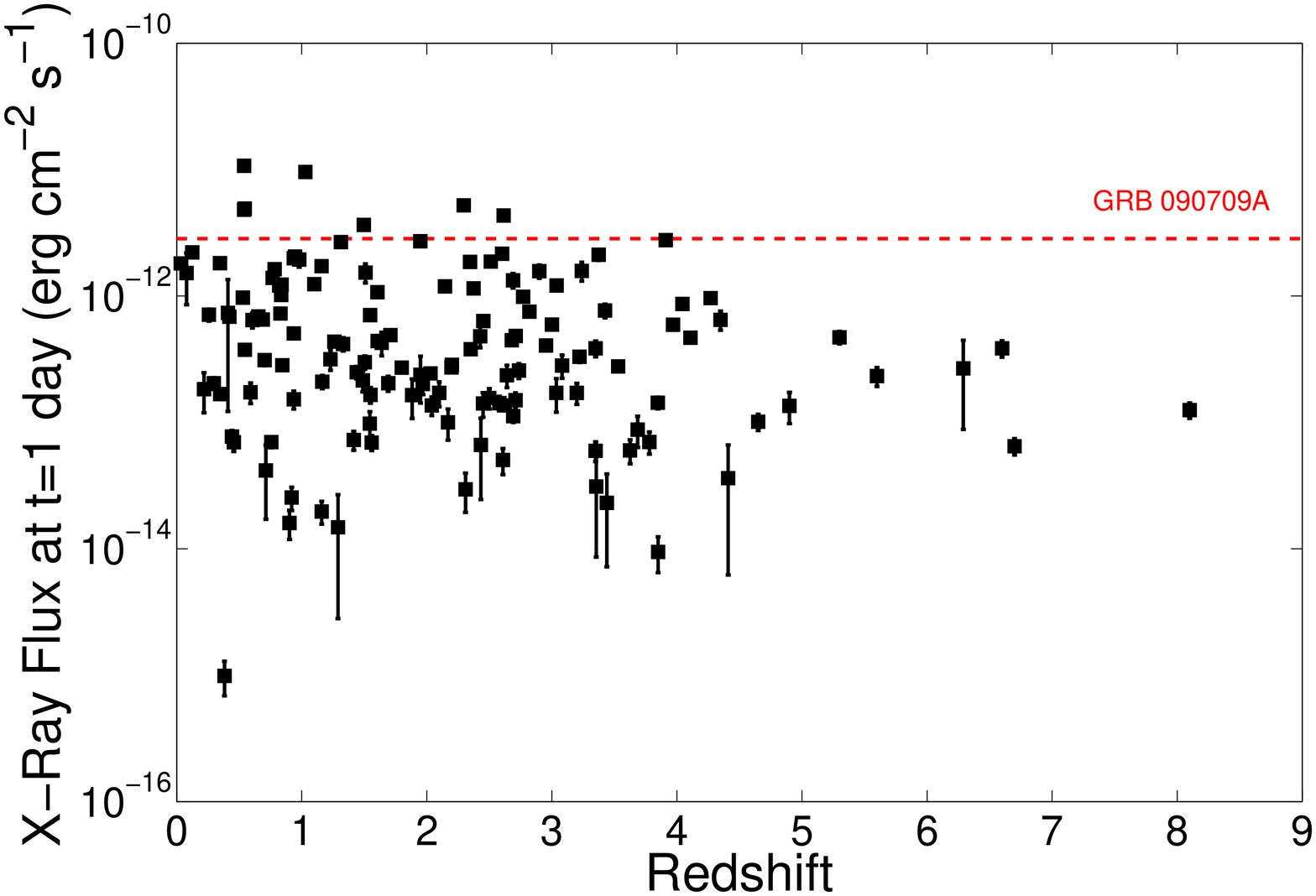}{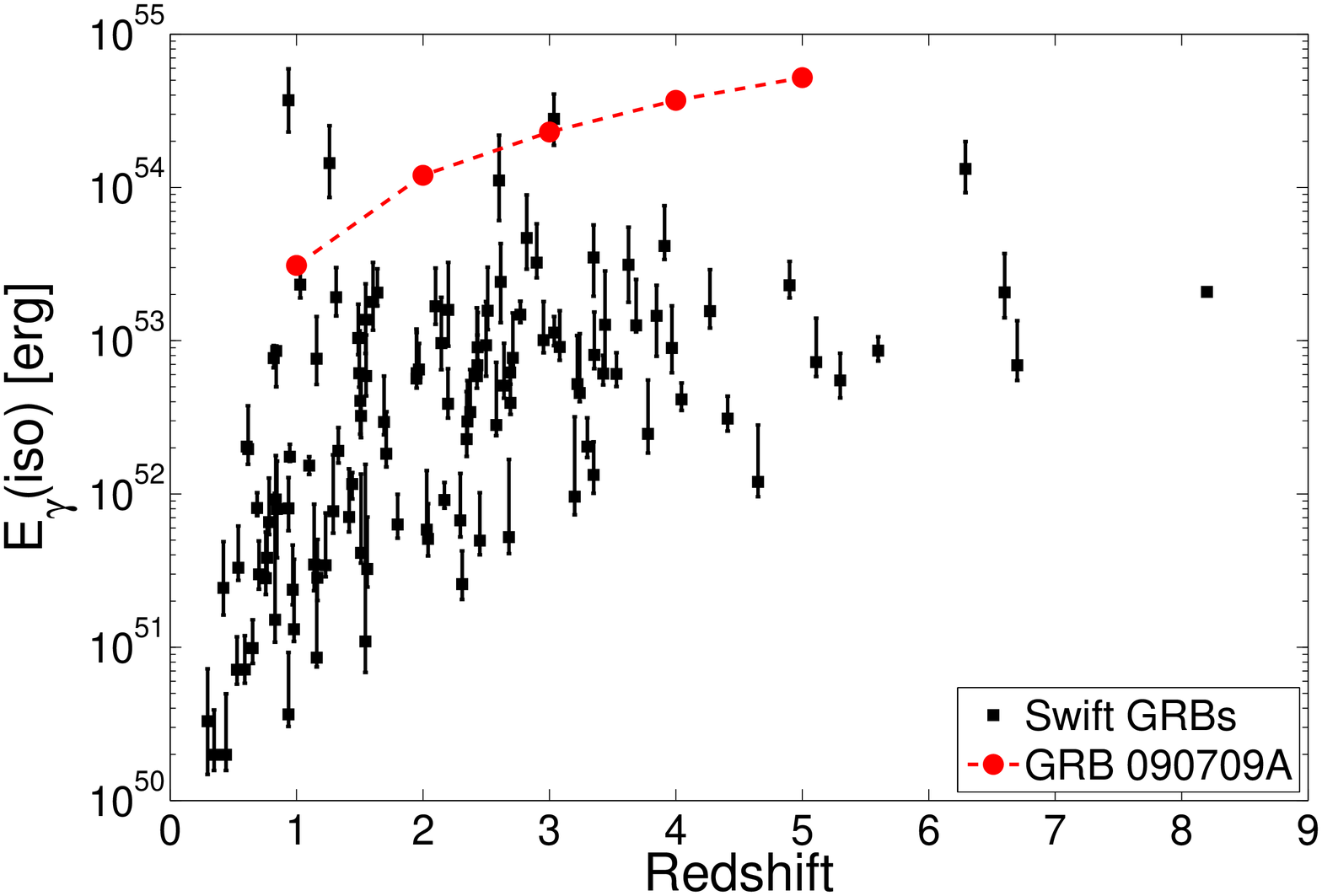}}
  \caption{\textit{Left:} The 0.3--10\,keV flux at a common time of $t=1$\,day (observer frame)
                 for all \swift\ GRBs with known redshift.  GRB\,090709A is indicated with a dashed
                 horizontal line.  \textit{Right:} The prompt gamma-ray (1--10$^{4}$\,keV rest frame)
                 energy release for all \swift\ GRBs with known redshift.  GRB\,090709A is shown
                 for $z = 1$, 2, 3, 4, and 5.  In both cases, GRB\,090709A is one of the brightest
                 events in the \swift\ era, suggesting an approximate upper limit on the GRB 
                 redshift of $z \lesssim 3$. }
\label{fig:redshift}
\end{figure*}

%%%%%%%%%%%%%%%%%%%%%%%%%%%%%%%%%%%%%%%%%%%%%%%%%%%%%%%%%%%%%%%%%%%%%%%%

To do so, we require some knowledge of the distance to GRB\,090709A.
\citet{gnv+07} have demonstrated that the detection of X-ray absorption in
excess of the Galactic value can be used to place a probabilistic upper limit 
on the distance to GRBs; the steep redshift dependence of soft X-ray 
absorption ($\sim (1+z)^{8/3}$) effectively precludes the observation of distant
events with large column densities.  Using Equation 1 from \citet{gnv+07} 
and our derived X-ray column excess, we find $z \lesssim 3.5$.  We caution,
however, that some well-known events are known to violate this redshift
limit (e.g., GRB\,080607; \citealt{psp+09}).  

A more strict limit can be placed using the P60 and Faulkes Telescope 
\citep{GCN.9648} $r^{\prime}$-band afterglow detections.  If we assume the 
Ly-$\alpha$ break in the GRB rest frame falls at the red edge of the
observed $R$-band filter, we conclude $z \lesssim 4.5$.

An alternate approach is to use the observed BAT and XRT luminosity
functions as a guide.  The left panel of Figure~\ref{fig:redshift} displays
the X-ray afterglow flux (0.3--10\,keV) at a common time of $t = 1$\,day
(observer frame) for all \swift\ GRBs to date with known redshifts.  The X-ray
afterglow of GRB\,090709A is one of the brightest in the \swift\ era,
well above any previous event with $z > 4$.

Much like the X-ray flux, the BAT fluence from GRB\,090709A is also
one of the largest ever seen by \swift.  In the right panel of 
Figure~\ref{fig:redshift}, we plot the rest frame 1--10$^{4}$\,keV 
isotropic energy release ($E_{\gamma,\mathrm{iso}}$) for all BAT GRBs with
known redshifts, along with K-corrected values for GRB\,090709A
at $z$ = 1, 2, 3, 4, and 5.  To be consistent with the observed BAT
distribution, GRB\,090709A should fall at $z \lesssim 3$.  More
quantitatively, applying the GRB world model from \citet{bbp09},
we find $z < 2.5$ ($z < 1$) at 99\% (90\%) confidence.

An approximate lower limit can be derived using the lack of a host candidate
at the location of the afterglow of GRB\,090709A.  With only a handful of
exceptions (e.g., \citealt{cfp+08}), the observed host-offset distribution of 
long-duration GRBs is limited to $r \lesssim 10$\,kpc \citep{bkd02,fls+06}.
Long-duration GRB hosts are typically blue, irregular galaxies
(e.g., \citealt{fls+06,wbp07}) with low metallicity (e.g., \citealt{sgb+06})
and large specific star-formation rates (e.g., \citealt{chg04}).
If we take the low-metallicity dwarf Small Magellanic Cloud as a proxy for
our GRB host ($M_{V} \approx -16.6$ mag), the non-detection in the 
$i^{\prime}$ band with the GTC requires $z \gtrsim 0.5$.

Combining these results, we believe the redshift of GRB\,090709A falls 
somewhere in the range $0.5 \lesssim z \lesssim 3.0$.  For 
comparison, \citet{dei+09} favor a somewhat larger distance 
($3.7 < z < 4.5$; 90\% confidence interval) based only on X-ray 
spectral fits assuming an intrinsic power-law spectrum.  
The distance limits derived here are not particularly
constraining.  In what follows, we
adopt a fiducial redshift of $\sim 1$.  The implied isotropic gamma-ray 
energy release is then $E_{\gamma,\mathrm{iso}} \approx 3.1 \times 10^{53}$\,erg,
and the observed host-galaxy limits correspond to $M_{i^{\prime}} 
(\mathrm{AB}) > -18.6$\,mag and $M_{K} (\mathrm{AB}) > -19.5$\,mag.  Both
results are consistent with the observed long-duration GRB population
\citep{bkb+07,sgl09}.  Where appropriate, we discuss the impact of varying the 
distance on the results.

Clearly the X-ray afterglow provides the best sampling, so we begin our
investigation there.  After the first $\sim 10^{3}$\,s, the X-ray decay
is quite smooth.  There is no evidence for dramatic variability on short
time scales seen, for example, in the X-ray light curve of GRB\,070610
\citep{kck+08,sks+08,cdg+08}.  Most strikingly, a detailed timing 
analysis of X-ray afterglow of both XRT and \textit{XMM-Newton}
observations of GRB\,090709A reveals no evidence for periodicity 
superposed on the overall power-law decline \citep{GCN.9696,dei+09}.  
Such behavior is almost always seen in the decaying X-ray phase of SGR flares, 
and its absence is a challenge to any magnetar model.

On the other hand, synchrotron afterglow theory predicts a series
of possible relationships between the spectral and temporal indices
known as closure relations.  For the X-ray afterglow of GRB\,090709A,
the observed values agree well with theoretical predictions if the
X-ray band falls below the cooling frequency (i.e., $\nu_{X} < 
\nu_{c}$) and the shock wave is expanding into a constant-density
circumburst medium ($\rho \propto r^{0}$): $\alpha = 3 \beta / 2$.
It is relatively unusual to observe the cooling frequency above
the X-ray bands at this early time, though not
unprecedented (e.g., GRB\,060418; \citealt{cfh+09}).  
We note, however, that the case of the X-ray band falling above 
the cooling frequency ($\nu_{X} > \nu_{c}$) is marginally acceptable, 
particularly if the shallower initial decay index from the broken 
power-law fit is used (light curve behavior above $\nu_{c}$ is
independent of the circumburst medium density profile).  
The case of the X-ray bandpass falling
below $\nu_{c}$ in a wind-like circumburst environment ($\rho
\propto r^{-2}$) is strongly ruled out for both X-ray light curve
fits.  

The observed optical spectral index ($\beta_{O} = 3.8$) is much 
steeper than predicted by afterglow theory.  Furthermore, if we 
consider the last reported $K$-band detection from Calar Alto
at $t \approx 0.8$\,day (when the X-ray light curve is dominated 
by afterglow light), the measured X-ray to optical spectral 
index is $\beta_{OX} \approx 0.26$, well below the ``dark'' burst 
threshold ($\beta_{OX} < 0.5$; \citealt{jhf+04}).  Both facts suggest
a significant amount of optical extinction along the line of sight.

In order to make the intrinsic $K$-band to X-ray flux ratio consistent
with predicted values ($\beta_{OX} > 0.5$), the observed $K$-band flux
must be suppressed by dust along the line of sight by at least
$A_{K,\mathrm{obs}} \gtrsim 2$\,mag.  However, if the synchrotron 
cooling frequency does indeed fall above the X-ray bandpass, the 
SED should be continuous from the X-rays through the optical;
i.e., $\beta_{OX} \approx 1$.  In this case, the required host
extinction would rise sharply, to $A_{K,\mathrm{obs}} \approx 6$\,mag.

We can estimate the rest frame $V$-band extinction 
($A_{V,\mathrm{host}}$) by assuming the optical SED remains largely
unchanged from our measurement at $t_{0} + 350$\,s ($\beta_{O}
\approx 3.8$).  If the optical to X-ray spectral index varies
from $0.5 \lesssim \beta_{OX} \lesssim 1.0$, we find a host
galaxy extinction of $A_{V,\mathrm{host}} \approx 4$--8\,mag
($A_{V,\mathrm{host}} \approx 2$--6\,mag) for $z \approx 1$
($z \approx 3$).

Though most well-studied long-duration afterglow sightlines 
exhibit relatively small amounts of host extinction 
($\langle A_{V} \rangle 
\approx 0.2$\,mag; \citealt{kkz06,kkz+07,spo+09}), this result is likely 
strongly
biased toward low-$A_{V}$ sightlines (e.g., \citealt{ckh+09}).  GRB
afterglows with host $A_{V}$ values as large as $\sim 3$\,mag  
have been reported in the literature (e.g., \citealt{psp+09,rvw+07}),
and even larger values have been inferred for GRBs without measured
redshifts (e.g., \citealt{tlr+08,pcb+09}).  Like GRB\,090709A, these
highly obscured events often masquerade
as high-redshift candidates at first (e.g., \citealt{GCN.9634,GCN.9635}).
Large host columns are on occasion expected given the likely 
origin of long-duration GRBs in giant molecular clouds in the
disks of their host galaxies.  Furthermore, the non-detection
of the host galaxy at mm wavelengths \citep{GCN.9685} is not 
particularly constraining, as most dark-burst host galaxies do not 
differ dramatically from the larger long-duration GRB population,
suggesting a patchy dust distribution
(e.g., \citealt{bck+03,pcb+09}).

Examining a variety of different extinction laws over the 
redshift range $0.5 \lesssim z \lesssim 3$, we find that a large
selective extinction coefficient ($R_{V} \approx 4$--15) is required
to recover an intrinsic optical spectral slope of  $0 \lesssim 
\beta_{O} \lesssim 1$ in nearly all cases.  
For comparison, the global value of $R_{V}$ 
for the Milky Way (SMC) is 3.08 (2.93; \citealt{p92}).  A large 
value of $R_{V}$ ($\sim 4$) was inferred for the heavily 
obscured GRB\,080607 \citep{psp+09}, similar to the lines of 
sight toward molecular clouds in the Milky Way, and may therefore
be common for heavily extinguished environments.  The extinction
law from \citet{cks94} can provide a reasonable fit across the
entire redshift range of interest.  To a large extent this is due to 
the featureless nature of this extinction law outside the far
ultraviolet, which by default maintains the power-law shape
of the observed optical SED.  On the other hand,
an SMC-like extinction law is only consistent with $z \gtrsim 
1.5$.  Extinction laws with a strong 2175\,\AA\ bump 
(Milky Way, LMC) also favor redshifts near our upper limit.  

We have also attempted to constrain the host galaxy column
density ($N_{H,\mathrm{host}}$) using the late-time ($t \gtrsim
10^{3}$\,s) XRT spectrum.  Not surprisingly, the required column
rises sharply as a function of redshift, from $N_{H,\mathrm{host}}
= (3.5 \pm 0.5) \times 10^{21}$\,cm$^{-2}$ at $z \approx 0.5$
to $N_{H,\mathrm{host}} = (3.6 \pm 0.5) \times 10^{22}$\,cm$^{-2}$
at $z \approx 3$.  Our X-ray spectral fits are consistent with
those derived in \citet{dei+09} given somewhat lower
preferred redshifts.  At the smallest distances, the inferred 
dust-to-metals ratio ($A_{V,\mathrm{host}} / N_{H,\mathrm{host}}
\approx 1$--2 $\times 10^{-21}$) is in excess of the Galactic
value.  This result is inconsistent with observations of 
previous GRB sightlines (e.g., \citealt{gw01,wfl+06}), 
where the dust-to-metals ratio is typically well below the Galactic
value.  At larger distances, $A_{V,\mathrm{host}}$ drops and 
$N_{H,\mathrm{host}}$ rises, lowering the inferred ratio to 
6--17 $\times 10^{-23}$ at $z \approx 3$.  This result
suggests GRB\,090709A may fall at the high end of our
preferred distance scale.

\begin{figure*}[t!]
  \centerline{\plotone{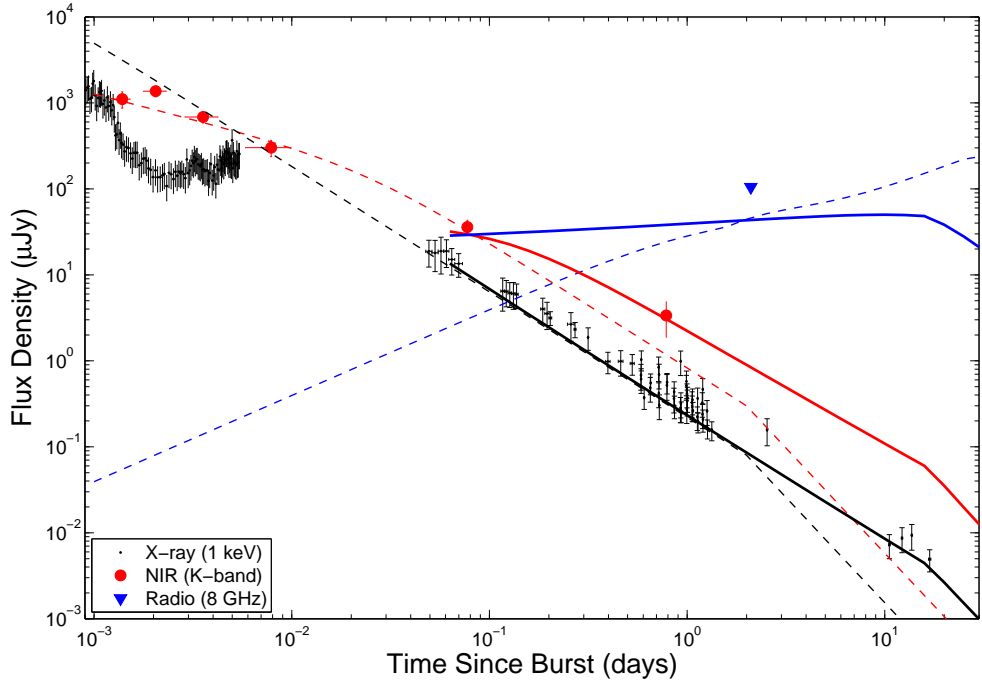}}
  \caption{The X-ray, $K$-band, and radio afterglow of GRB\,090709A.  The dashed lines
                 indicate a model that attempts to explain the entirety of the observations
                 (the model has the following parameters: $E_{\mathrm{KE,iso}} = 1.5 \times
                 10^{53}$\,erg, $\theta = 0.2$, $p=2.01$, $\epsilon_{e} = 0.5$, $\epsilon_{B} =
                 0.2$, $z=1$, $n=40$\,cm$^{-3}$, $A_{K,\mathrm{obs}} = 6$\,mag).  While this model can
                 account for most of the observed features, it implies an extremely small
                 cooling frequency ($\nu_{c} < \nu_{O}$ for $t < 0.01$\,day).  The solid lines
                 shows a model with more reasonable physical parameters designed to
                 explain the late-time ($t \gtrsim 0.1$\,day) data (the model has the
                 following parameters: $E_{\mathrm{KE,iso}} = 1.5 \times
                 10^{53}$\,erg, $\theta = 0.1$, $p=2.06$, $\epsilon_{e} = 0.6$, $\epsilon_{B} =
                 0.1$, $z=1$, $n=0.01$\,cm$^{-3}$, $A_{K,\mathrm{obs}} = 3$\,mag). The lack of optical
                 and radio coverage at these times makes such models highly degenerate.}
\label{fig:aglow}
\end{figure*}

Assuming a \citet{cks94} extinction law with a large $R_{V}$ can
account for the optical SED, we have constructed sample
models for the X-ray, $K$-band, and radio afterglow emission from 
GRB\,090709A using the software described by \citet{yhs+03}.
No single forward-shock model can account for all of the observed
data; a sample attempt is shown with dashed lines in 
Figure~\ref{fig:aglow}.  The parameters of the dashed-line model
are relatively contrived to provide an extremely small cooling
frequency ($\nu_{c} < \nu_{O}$ for $t < 0.01$\,day), unlike what we
inferred previously based on the X-ray spectral and temporal 
indices.  This is not entirely surprising, as we know the early
X-ray emission is often dominated by other components besides
the afterglow \citep{nkg+06}.  In particular, energy injection
in this phase can significantly alter the subsequent evolution
of the afterglow.  

%%%%%%%%%%%%%%%%%%%%%%%%%%%%%%%%%%%%%%%%%%%%%%%%%%%%%%%%%%%%%%%%%%%%%%

%%%%%%%%%%%%%%%%%%%%%%%%%%%%%%%%%%%%%%%%%%%%%%%%%%%%%%%%%%%%%%%%%%%%%%%%

If we only consider the emission at $t \gtrsim 0.1$\,day, when the
afterglow emission from the forward shock appears to dominate
the observed emission, we can easily find good models with
$\nu_{c} > \nu_{X}$.  One example is shown as the solid line
in Figure~\ref{fig:aglow}.  Lower redshifts typically provide better
fits to the bright X-ray flux.  However, the lack of optical and
radio data at these times leaves the basic physical parameters
of such models largely unconstrained.

%%%%%%%%%%%%%%%%%%%%%%%%%%%%%%%%%%%%%%%%%%%%%%%%%%%%%%%%%%%%%%%%%%%%%%%%%%%%

\section{Conclusion}
\label{sec:discussion}
We now return again to our original question: what is the origin of GRB\,090709A?
A careful examination of the high-energy light curve indicates that the previously
claimed detection of quasi-periodic oscillations was 
in our opinion overstated: with a 
proper accounting for the underlying spectrum, we detect the periodic
signal with only marginal ($\sim 2\sigma$) significance.  Together with the
lack of obvious environmental clues of recent star formation, it seems
unlikely that GRB\,090709A could be caused by an SGR-like outburst
from a highly magnetized neutron star.

Independent of the high-energy properties, we have shown that
GRB\,090709A was almost certainly a cosmologically distant
event.  Even if associated with an older stellar population and 
therefore free of the requirement to be nearby recent star formation,
the X-ray and optical extinction in excess of the Galactic value
are difficult to reconcile with an origin in the Milky Way or 
nearby universe.  The bright high-energy fluence and X-ray 
afterglow favor a more nearby event ($z \approx 1$), while the
X-ray and optical extinction properties point to a more distant
origin ($z \approx 3$).  Perhaps the true value falls somewhere
in between.

A long-duration GRB (and, by assumption, the core collapse of a
massive star), on the other hand, can naturally account
for the majority of the observed properties of GRB\,090709A: the large
line-of-sight extinction, the lack of an obvious host galaxy, and the 
late-time afterglow decay.  In many ways GRB\,090709A appears to
be a relatively close analog of GRB\,080607 \citep{psp+09}: 
in particular the large host extinction ($A_{K,\mathrm{obs}}
\gtrsim 2$\,mag) and the large selective extinction ($R_{V} \gtrsim
4$) are shared by both events.  Such highly obscured GRBs
may be relatively common, but are more likely to be discovered recently
due to the prompt localization capabilities of \swift\ and
rapid ground-based follow-up in the NIR.  If so, it will be
important to incorporate such optically dark events into systematic
studies of the GRB population to minimize observational
bias (e.g., \citealt{pcb+09}).

Subsequent observations could definitively resolve this issue:
additional outbursts from this location would require an SGR-like
(i.e., non-destructive) progenitor.  On the other hand, the detection
of a faint host galaxy at the location of the afterglow (and in 
particular measurement of a substantial redshift) would provide even stronger
evidence in support of our long-duration GRB hypothesis.

%%%%%%%%%%%%%%%%%%%%%%%%%%%%%%%%%%%%%%%%%%%%%%%%%%%%%%%%%%%%%%%%%%%%%%%%%%%%%

\acknowledgements
We wish to thank Eliot Quataert and Tony Piro for valuable discussions, Cullen Blake
and Bethany Cobb for their assistance in automating and operating the PAIRITEL 
telescope, C.~\'Alvarez for his help with the GTC observations, G.~Bergond for his
support during the 3.5\,m CAHA observations, Mansi Kasliwal and Fiona Harrison
for assistance in operating P60, and Andy Boden for 
aquiring the Keck/LGS data.  S.B.C.~and A.V.F.~are grateful for generous support 
from Gary and Cynthia Bengier, the Richard and Rhoda Goldman Fund, and National 
Science Foundation (NSF) grant AST–0908886.  A.N.M.~acknowledges support from an NSF 
Graduate Research Fellowship.  The research  of J.G.~and A.J.C.-T.~are supported 
by the Spanish programs ESP2005-07714-C03-03, AYA2004-01515,   
AYA2007-67627-C03-03, AYA2008-03467/ESP, and AYA2009-14000-C03-01.  
P60 operations are funded in part by NASA through the \textit{Swift} 
Guest Investigator Program (grant number NNG06GH61G).  This publication has
made use of data obtained from the {\it Swift} interface of the High-Energy Astrophysics Archive
(HEASARC), provided by NASA’s Goddard Space Flight Center.  PAIRITEL is operated by the 
Smithsonian Astrophysical Observatory
(SAO) and was made possible by a grant from the Harvard University Milton Fund,
a camera loan from the University of Virginia, and continued support of the SAO and UC
Berkeley. The PAIRITEL project and those working on PAIRITEL data are further 
supported by NASA/{\it Swift} Guest Investigator
grant NNG06GH50G and NNX08AN84G. Based in part on observations collected at the Centro
Astron\'omico Hispano Alem\'an  (CAHA) at Calar Alto, operated jointly by the 
Max-Planck Institut  f\"ur Astronomie and the Instituto de Astrof\'isica de
Andaluc\'ia (CSIC).  We thank Calar Alto Observatory for allocation of Director's 
Discretionary Time.  Some of the data presented herein were obtained
at the W. M. Keck Observatory, which is operated as a scientific partnership among the California
Institute of Technology, the University of California, and NASA; the Observatory
was made possible by the generous financial support of the W. M. Keck Foundation. The
authors wish to recognize and acknowledge the very significant cultural role and reverence
that the summit of Mauna Kea has always had within the indigenous Hawaiian community;
we are most fortunate to have the opportunity to conduct observations from this mountain.

%%%%%%%%%%%%%%%%%%%%%%%%%%%%%%%%%%%%%%%%%%%%%%%%%%%%%%%%%%%%%%%%%%%%%%%%%%%%%

{\it Facilities:} \facility{Swift (BAT; XRT)}, \facility{FLWO:1.3m (PAIRITEL)}, \facility{PO:1.5m}, 
\facility{CAO:3.5m (OMEGA$_{2000}$)}, \facility{GTC (OSIRIS)}, \facility{Keck:II (NIRC2/LGS)}, 
\facility{PO:1.2m (QUEST)}, \facility{VLA}

%%%%%%%%%%%%%%%%%%%%%%%%%%%%%%%%%%%%%%%%%%%%%%%%%%%%%%%%%%%%%%%%%%%%%%%%%%%%%

%%%%%%%%%%%%%%%%%%%%%%%%%%%%%%%%%%%%%%%%%%%%%%%%%%%%%%%%%%%%%%%%%%%%%%%%%%%%%

\end{document}